\documentclass{PoS}

\usepackage{amssymb, amsbsy, amsfonts}
\usepackage{amsmath, amsthm, cite}

\allowdisplaybreaks[3]

\let\set\mathbb
\newcommand{\ep}{\varepsilon}{

\newcommand{\SigmaP}{\texttt{Sigma}}

\newcounter{mmacnt}
\def\restartmma{\setcounter{mmacnt}{0}}
\restartmma \catcode`|=\active
\def|#1|{\mathrm{#1}}
\catcode`|=12
\newenvironment{mma}{
 \par\smallskip
 \catcode`|=\active
 \parskip=0pt\parindent=0pt 
 \small
 \def\In##1\\{%
   \def\linebreak{\hfill\break\null\qquad}%
   \refstepcounter{mmacnt}
   \hangindent=2.5em\hangafter=0
   \leavevmode
   \llap{\tiny\sffamily In[\arabic{mmacnt}]:=\kern.5em}%
   \mathversion{bold}\footnotesize$\tt\bf\displaystyle##1$\normalsize
   \mathversion{normal}\par
 }%
 \def\Print##1\\{%
   \def\linebreak{\hfill\break}%
   \hangindent=2.5em\hangafter=0
   \leavevmode\footnotesize ##1\par}%
 \def\Out##1\\{%
   \def\linebreak{$\hfill\break\null\hfill$}%
   \kern\abovedisplayskip\par
   \hangindent=2.5em\hangafter=0
   \leavevmode
   \llap{\tiny\sffamily Out[\arabic{mmacnt}]=\kern.5em}
   \footnotesize$\displaystyle\tt##1$\normalsize\hfill\null\par
   \kern\belowdisplayskip
 }%
 \def\Warning##1##2\\{%
   \def\linebreak{\hfill\break}%
   \hangindent=2.5em\hangafter=0
   \leavevmode
   {\scriptsize##1 : ##2}\par}%
}{%
 \par\smallskip
}

\title{{\footnotesize DESY 12-164, DO-TH-12/30 SFB/CPP-12-73, LPN12-104}\\
Advanced Computer Algebra Algorithms for the Expansion of Feynman Integrals}

\ShortTitle{Advanced Computer Algebra Algorithms for the Expansion of Feynman Integrals}

\author{Jakob Ablinger\\
        Research Institute for Symbolic Computation (RISC)\\
        Johannes Kepler University Linz\\
        Altenberger Str. 69, 4040 Linz, Austria
        E-mail: \email{Jakob.Ablinger@risc.jku.at}}

\author{Johannes Bl\"umlein\\
        Deutsches Elektronen--Synchrotron (DESY), Zeuthen\\ 
        Planetenalle 6, D-15735 Zeuthen, Germany
        E-mail: \email{Johannes.Bluemlein@desy.de}}

\author{Mark Round\\
        Research Institute for Symbolic Computation (RISC)\\
        Johannes Kepler University Linz\\
        Altenberger Str. 69, 4040 Linz, Austria
        E-mail: \email{Mark.Round@risc.jku.at}}

\author{\speaker{Carsten Schneider}\\
        Research Institute for Symbolic Computation (RISC)\\
        Johannes Kepler University Linz\\
        Altenberger Str. 69, 4040 Linz, Austria
        E-mail: \email{Carsten.Schneider@risc.jku.at}}

\abstract{Two-point Feynman parameter integrals, with at most one mass and containing local operator insertions in $4+\ep$-dimensional Minkowski space,
can be transformed to multi-integrals or multi-sums over hyperexponential and/or hypergeometric functions depending on a discrete parameter $n$. Given such a specific representation, we utilize an enhanced version of the multivariate Almkvist--Zeilberger algorithm (for multi-integrals) and a common summation framework of the holonomic and difference field approach (for multi-sums) to calculate recurrence relations in $n$. Finally, solving the recurrence we can decide efficiently if the first coefficients of the Laurent series expansion of a given Feynman integral can be expressed in terms of indefinite nested sums and products; if yes, the all $n$ solution is returned in compact representations, i.e., no algebraic relations exist among the occurring sums and products.}

\FullConference{Loops and Legs in Quantum Field Theory - 11th DESY Workshop on Elementary Particle Physics,\\
		April 15-20, 2012\\
		Wernigerode, Germany}

\begin{document}

\section{Introduction}

We consider
Feynman integrals in $D$-dimensional Minkowski space with one time- and $(D-1)$
Euclidean space dimensions, $\varepsilon = D - 4$ and $\varepsilon \in {\mathbb R}$ with
$|\varepsilon| \ll 1$, and with at most one mass.
Here the discrete Mellin parameter $n$ comes from local operator insertions. 
As worked out in detail in~\cite{Blumlein:2009ta,BKSF:12} these integrals can be 
transformed
to integrals of the form\begin{eqnarray}
\label{Equ:HypInt}
{\cal I}(\ep,n) = C(\ep, n, M) \int_0^1 dx_1 \ldots \int_0^1 dx_m
\frac{\sum_{i=1}^k \prod_{{l}=1}^{r_i}
[P_{i,l}(x_1,\dots,x_m)]^{\alpha_{i,l}(\varepsilon,n)}}{[Q(x_1,\dots,x_m)]^{\beta(\varepsilon)}}~,
\end{eqnarray}
with $k\in\set N$, $r_1,\dots,r_k\in\set N$ and
where $\beta(\ep)$ is  given by a rational function in $\ep$, i.e., $\beta(\ep)\in\set Q(\ep)$, and similarly
$\alpha_{i,l}(\ep,n) = n_{i,l} n + \overline{\alpha}_{i,l}$ for some $n_{i,l} \in \{0,1\}$ and $\overline{\alpha}_{i,l}\in\set Q(\ep)$, see also \cite{BOGNER}
in the case no local operator insertions are present.
$C(\ep, n, M)$ is a factor, which depends on the dimensional parameter $\ep$,
the integer parameter $n$ and the mass $M$. $P_i(x_1,\dots,x_m), Q(x_1,\dots,x_m)$ 
are polynomials in the $x_i$. Integrals of the type (\ref{Equ:HypInt}) emerge in the 
calculation of unpolarized and polarized massive operator matrix elements (OMEs)
\cite{Bierenbaum:2007qe,Bierenbaum:2007dm,BBKS:08,Bierenbaum:2009zt,
Bierenbaum:2009mv,Blumlein:2009rg,HYP2,ABHKS:12,BHKC:13} and in 
other single scale higher loop calculations. In 
\cite{Bierenbaum:2009mv,Blumlein:2009rg} 3-loop moments of the corresponding OMEs 
have been calculated.

In addition, such integrals~\eqref{Equ:HypInt} can be transformed to proper hypergeometric multi-sums of the form\footnote{For convenience, we assume that the summand is written terms of the Gamma function $\Gamma(x)$. Later, also Pochhammer symbols or binomial coefficients are used which can (if necessary) be rewritten in terms of Gamma-functions.}
\begin{equation}\label{Eq:GenericMultiSum}
{\cal S}(\ep,n) = \sum_{n_1=1}^\infty ... \sum_{n_r=1}^\infty
\sum_{k_1=1}^{L_1(n)} ... \sum_{k_v=1}^{L_v(n,k_1, ..., k_{v-1})}
\sum_{k=1}^l C_k(\ep, n, M)
\frac{\Gamma(t_{1,k}) \ldots \Gamma(t_{v',k})}	
     {\Gamma(t_{v'+1,k}) \ldots \Gamma(t_{w',k})}.
\end{equation}
Here the upper bounds $L_1(n),\dots,L_{v}(n,k_1,\dots,k_{v-1})$ are integer linear (i.e., linear combinations of the variables over the integers) in
the dependent parameters or $\infty$, and $t_{l,k}$ are linear combinations of the $n_1,\dots,n_r$, of the $k_1,\dots,k_v$, and of $\ep$ over $\set Q$.

Finally, if the sums~\eqref{Eq:GenericMultiSum} are uniformly convergent, one of the most common tactics is as follows. First one expands the summand of~\eqref{Eq:GenericMultiSum}, say
\begin{equation}\label{Equ:SummandExpand}
F(n,n_1,\dots,n_r,v_1,\dots,v_k)=F_t(n,n_1,\dots,v_k)\ep^t+F_t(n,n_1,\dots,v_k)\ep^{t+1}+\dots
\end{equation}
with $t\in\set Z$
by using formulas such as
\begin{equation}\label{Eq:GammaExpand}
\Gamma(n+1+\bar{\ep}) = \frac{\Gamma(n) \Gamma(1+\bar{\ep})}{B(n,1+\bar{\ep})}\text{ and }B(n, 1 + \bar{\ep}) = \frac{1}{n}\exp\left(\sum_{k=1}^\infty \frac{(-\bar{\ep})^k}{k} S_k(n)\right)
= \frac{1}{n}\sum_{k=0}^\infty (-\bar{\ep})^k S_{\underbrace{\mbox{\scriptsize 1, \ldots
,1}}_{\mbox{\scriptsize
$k$}}}(n)
\end{equation}
with $\bar{\ep} = r \ep$ for some $r\in\set Q$.
Here 
$B(x,y)=\Gamma(x)\Gamma(y)/\Gamma(x+y)$ denotes the Beta-function and $S_{\mbox{\scriptsize 1, \ldots
,1}}(n)$ is a special instance of the harmonic sums~\cite{Vermaseren:99,Bluemlein:99} defined by
\begin{equation}\label{Equ:HarmonicSums}
S_{m_1,\dots,m_k}(n)=
\sum_{i_1=1}^n\frac{\text{\small$\text{sign}(m_1)^{i_1}$}}{i_1^{|m_1|}}\dots
\sum_{i_k=1}^{i_{k-1}}\frac{\text{\small$\text{sign}(m_k)^{i_k}$}}{i_k^{|m_k|}}
\end{equation}
with $m_1,\dots,m_k$ being nonzero integers. Then one applies the summation signs to each of the coefficients in~\eqref{Equ:SummandExpand}. I.e., the $i$th coefficient of the $\ep$-expansion of~\eqref{Eq:GenericMultiSum} yields 
$$\sum_{n_1=1}^\infty ... \sum_{n_r=1}^\infty
\sum_{k_1=1}^{L_1(n)} ... \sum_{k_v=1}^{L_v(n,k_1, ..., k_{v-1})}
\sum_{k=1}^lF_i(n,n_1,\dots,n_r,v_1,\dots,v_k).$$
Then the essential problem is the simplification of these sums to special functions, like, e.g., harmonic sums, $S$--sums~\cite{Moch:02}
\begin{equation}\label{Equ:SSums}
S_{m_1,\dots,m_k}(x_1,\dots,x_k,n)=
\sum_{i_1=1}^n\frac{x_1^{i_1}}{i_1^{m_1}}\dots
\sum_{i_k=1}^{i_{k-1}}\frac{x_k^{i_k}}{i_k^{m_k}},
\end{equation}
cyclotomic harmonic sums~\cite{ABC:11}, or more generally to indefinite nested sums and products~\cite{Schneider:10b}.
For various special cases, this simplification can be carried out with efficient 
methods available, e.g., in {\tt Form}; see in~\cite{Vermaseren:99,Moch:02}. 

\noindent More  general sums can be handled with the Mathematica package \texttt{EvaluateMultiSum} \cite{SchneiderSummation:10,BHC11} based on the summation package \texttt{Sigma} \cite{Schneider:07a}. With the underlying difference field algorithms~\cite{Karr:81,Schneider:05a,Schneider:05c,Schneider:08c,Schneider:10b,ABPS:12} generalizing the hypergeometric summation paradigms~\cite{AequalB} to multi-sum\-mation we are currently simplifying sums up to nesting depth 7. The compact representation of the output, i.e., the elimination of all algebraic relations among the arising indefinite nested sums and products can be guaranteed by difference field theory~\cite{Schneider:10c}. For harmonic sums, cyclotomic sums, $S$-sums and cyclotomic $S$-sums and their infinite versions (quasi-)shuffle algebras are utilized; see e.g.,~\cite{Bluemlein:04,ANCONT3,MZV:10,ABC:11,ABC:12} and references therein. In this regard, the Mathematica package \texttt{HarmonicSums} is heavily used~\cite{Ablinger:12}. This general machinery has been applied to non-trivial massive 3-loop diagrams arising, e.g., in~\cite{BBKS:08,HYP2,ABHKS:12,BHKC:13}.

\noindent Another possibility is the method of hyperlogarithms~\cite{Brown:09} which can be used to evaluate integrals of the form~\eqref{Equ:HypInt} for specific values $n\in\set N$ if one can set $\ep=0$. An adaption of this method for symbolic $n$ has been described and applied to massive 3-loop ladder graphs in~\cite{ABHKS:12}.

In this article we follow another promising approach for the all $n$ expansion\\
1. Calculate a recurrence in $n$ for the multi-integral~\eqref{Equ:HypInt} or multi-sum~\eqref{Eq:GenericMultiSum}\\
2. Given this recurrence and initial values (using, e.g., the \texttt{EvaluateMultiSum} package),\\ 
\hspace*{0.4cm}calculate the $\ep$-expansion by a recurrence solver for Laurent expansions.

\smallskip

\noindent In~\cite{BKSF:12} we followed this approach by calculating recurrences of multi-sums using techniques of~\cite{WZtheory} and efficient algorithms developed in~\cite{Wegschaider}.
Subsequently, we present two new techniques to compute such recurrences.
In the first approach we apply an enhanced and optimized version~\cite{Ablinger:12} of the multivariate Almkvist--Zeilberger algorithm~\cite{ZEILB} to calculate recurrences for Feynman integrals in the form~\eqref{Equ:HypInt}. In the second approach we use a common framework~\cite{Schneider:05b} within the summation package Sigma that combines difference field~\cite{Karr:81,Schneider:05a} and holonomic summation techniques~\cite{Zeilberger-holonomic,Chyzak} to compute recurrences for Feynman integrals in the form~\eqref{Eq:GenericMultiSum}.
Two new packages, \texttt{MultiIntegrate} by J. Ablinger and \texttt{RhoSum} by M. Round facilitate these tasks completely automatically for the integral or sum representation, respectively.

The outline of the article is as follows. In Section~\ref{Sec:LaurentSolver} we present the general idea of how the Laurent series representation can be calculated from the given recurrence representation. With this knowledge, we illustrate our  integration and summation methods in Sections~\ref{Sec:Integrate} and~\ref{Sec:Sum}, respectively.

\section{Finding Laurent series solutions of linear recurrences}\label{Sec:LaurentSolver}

One of the key ingredients of the summation and integration tools under consideration is a recurrence solver for $\ep$-expansions. 
To illustrate the ideas of the solver, we consider the single sum
\begin{equation}\label{Equ:SimpleSumProblem}
\begin{split}
\mathcal{S}(\ep, n)&=\sum_{k=1}^{\infty }\frac{B\left(\frac{\ep}{2}+k,n\right)}{(k+n)^
   2}=\sum_{k=1}^{\infty } \frac{\Gamma(n) \Gamma\big(\frac{\ep}{2}+k\big)}{(k+n)^2 \Gamma\big(\frac{\ep}{2}+k+n\big)}
\stackrel{?}{=}F_0(n)+F_1(n)\ep+F_2(n)\ep^2+\dots
\end{split}
\end{equation}
which is related, e.g., to sums arising in~\cite{BBKS:08}.
In order to derive the coefficients $F_i(n)$, we compute in a first step a recurrence relation using the summation package \texttt{Sigma}. Internally, it activates a difference field version of Zeilberger's creative telescoping paradigm~\cite{AequalB}. In our example it turns out that $\mathcal{S}(\ep, n)$ satisfies for all integer $n\geq1$, as an analytic function in $\varepsilon$ throughout an annular region centered by $0$,
the recurrence
\begin{multline}\label{Equ:SingleSumRec}
a_0(\varepsilon,n)\mathcal{S}(\ep, n)+a_1(\varepsilon,n) \mathcal{S}(\ep, n+1)+a_2(\varepsilon,n)\mathcal{S}(\ep, n+2)\\
=\frac{-4 (2 n+3) \left(\ep+4 n^2+12
   n+8\right) \Gamma
   \left(\frac{\ep}{2}+1\right)
   \Gamma (n+1)}{(n+1) (n+2)^2 (\ep+2
   n+2) \Gamma
   \left(\frac{\ep}{2}+n+1\right)}
\end{multline}
with
$a_0(\varepsilon,n)=-4 n (n+1)$,
$a_1(\varepsilon,n)=4 (n+1) (\ep-2n-3)$, and
$a_2(\varepsilon,n)=(\ep-2 n-4)^2$.
Next, we compute the $\ep$-expansion
$h_0(n)+h_1(n)\ep+h_2(n)\ep^2+\dots$
of the right hand side of~\eqref{Equ:SingleSumRec}
by formulas such as~\eqref{Eq:GammaExpand}; here we get
$h_0(n)=\tfrac{-8 (2 n+3)}{(n+1) (n+2)}$, $h_1(n)=\big(\tfrac{4(2 n+3) S_1(n)}{(n+1) (n+2)}\!+\!\tfrac{2 (2 n+3)^2}{(n+1)^2 (n+2)^2}\big)$,
$h_2(n)=-\big(\tfrac{(2 n+3)^2 S_1(n)}{(n+1)^2 (n+2)^2}\!+\!\frac{(2 n+3) S_1(n){}^2}{(n+1) (n+2)}+\tfrac{(2 n+3) S_2(n)}{(n+1) (n+2)}+\frac{(2 n+3)^2}{(n+1)^3 (n+2)^2}\big)$.
As a consequence, for the Laurent series expansion $\mathcal{S}(\ep,n)=F_0(n)+F_1(n)\ep+\dots$ the following relation holds:
\begin{equation}\label{Equ:SingleSumAnsatz}
\begin{split}
a_0(\ep,n)\Big[F_0(n)+F_1(n)\ep+\dots]
+a_1(\ep,n)\Big[F_0(n+1)+F_1(n+1)\ep+\dots]\\
+a_2(\ep,n)\Big[F_0(n+2)+F_1(n+2)\ep+\dots]
=h_0(n)+h_1(n)\ep+h_2(n)\ep^2+\dots
\end{split}
\end{equation}
Next, we expand the first two initial values\footnote{Here $\zeta_r$ stands for the Riemann Zeta function $\zeta(r)=\sum_{i=1}^{\infty}\frac{1}{i^r}$.} of ${\cal S}(\ep,n)$, $n=1,2$:
\begin{equation}\label{Equ:InitEpSolve}
\begin{split}
\mathcal{S}(\ep, 1)&=2-\zeta_2+(\tfrac{3}{2}-\zeta_2 )\ep+( -\tfrac{3 \zeta_2}{4}+\tfrac{\zeta_3}{4}+1)\ep^2+\dots,\\
\mathcal{S}(\ep,2)&=\tfrac{\zeta_2}{2}-\tfrac{3}{4}+(\frac{3 \zeta_2}{4}-\tfrac{41}{32})\ep+(\tfrac{21 \zeta_2}{32}-\frac{3 \zeta_3}{16}-\frac{53}{64})\ep^2+\dots
\end{split}
\end{equation}
by using the package \texttt{EvaluateMultiSum}~\cite{BHC11} in Mathematica; alternatively, one could use the package \texttt{Summer}~\cite{Vermaseren:99} in Form.
Then given the recurrence~\eqref{Equ:SingleSumAnsatz} and the first initial values~\eqref{Equ:InitEpSolve} (to be more precise the polynomials $a_i(\ep,n)$, the first coefficients $h_i(n)$, and the first values of their expansions), we are ready to calculate the first three coefficients of the all $n$ series expansion~\eqref{Equ:SimpleSumProblem}.

\noindent Namely, by setting $\varepsilon=0$ in~\eqref{Equ:SingleSumAnsatz}, it follows that the constant term $F_0(n)$ satisfies the recurrence
\begin{equation}\label{Equ:SingleSumConstrained}
a_0(0,n)F_0(n)+a_1(0,n)F_0(n+1)+a_2(0,n)F_0(n+2)=h_0(n).
\end{equation}
Note that together with $F_0(1)=2-\zeta_2$ and $F_0(2)=\tfrac{\zeta_2}{2}-\tfrac{3}{4}$ the sequence $F_0(n)$ is completely determined. At this point we exploit algorithms from~\cite{Petkov:92,Abramov:94,Schneider:05a,ABPS:12} which can constructively decide if a solution with certain initial values is expressible in terms of indefinite nested products and sums. More precisely, with~\SigmaP\ one obtains 
\begin{equation}\label{Equ:SingleSumF0}
F_0(n)=\frac{2 (-1)^n S_{-2}(n)}{n}+\frac{(-1)^n \zeta_2}{n}.
\end{equation}

\smallskip

\noindent Now, plugging in the partial solution
$$\mathcal{S}(\ep, n)=\frac{2 (-1)^n S_{-2}(n)}{n}+\frac{(-1)^n \zeta_2}{n}+F_1(n)\varepsilon+F_2(n)\varepsilon^2+\dots$$
into~\eqref{Equ:SingleSumAnsatz} and moving $F_0(n)$ to the right hand side yields 
\begin{multline}\label{Equ:RecWithEp}
a_0(\varepsilon,n)\big[F_1(n)\varepsilon+F_2(n)\varepsilon^2+\dots\big]+a_1(\varepsilon,n)\big[F_1(n+1)\varepsilon+F_2(n+1)\varepsilon^2+\dots\big]\\
+a_2(\varepsilon,n)\big[F_1(n+2)\varepsilon+F_2(n+2)\varepsilon^2+\dots\big]
=h'_0(n)\ep+h'_1(n)\ep^2+\dots
\end{multline}
with $h'_0(n)=\frac{4 (2 n+3) S_1(n)}{(n+1) (n+2)}+\frac{2 (4 n+5)}{(n+1)^2
(n+2)^2}$
and
\begin{align*}
h'_1(n)=\tfrac{-\big(2 n^2+9 n+8\big) (2 n+3)}{(n+1)^3 (n+2)^3}-\tfrac{(2 n+3)^2 S_1(n)}{(n+1)^2 (n+2)^2}-\tfrac{(2 n+3) S_1(n){}^2}{(n+1) (n+2)}+\tfrac{(-2 n-3) S_2(n)}{(n+1) (n+2)}+\tfrac{(-1)^n(2S_{-2}(n)+\zeta_2)}{n+2}.
\end{align*}
\normalsize
Observe that the coefficients and the inhomogeneous side of the recurrence~\eqref{Equ:RecWithEp} have all the factor $\ep$. Hence dividing the recurrence through $\ep$, we obtain again a recurrence of the form~\eqref{Equ:SingleSumAnsatz} where the explicitly given $h_i$ are changed to $h'_i$ and $F_1$ is the new constant term. In other words, we can repeat this construction process for $F_1(n)$. Namely, by coefficient comparison $F_1(n)$ is uniquely determined by
\begin{equation*}
a_0(0,n)F_1(n)+a_1(0,n)F_1(n+1)+a_2(0,n)F_1(n+2)=h'_0(n)
\end{equation*}
and the initial values given in~\eqref{Equ:InitEpSolve}.
Solving the recurrence with these initial values leads for all $n\geq1$ to the sum representation
\begin{equation}
F_1(n)=(-1)^n \big(\frac{5 S_{-3}(n)}{2 n}-\frac{3 S_{-2,1}(n)}{n}\big)+\frac{(-1)^n \zeta_2 S_1(n)}{n}+\frac{2 (-1)^n S_1(n) S_{-2}(n)}{n}.
\end{equation}
Similarly, one can loop further and calculates, e.g., the coefficients $F_0(n),\dots,F_6(n)$ (in terms of $\zeta_2$, $\zeta_3$, $\zeta_5$, $\zeta_7$ and harmonic sums up to weight 8) in about 30 minutes.

\medskip

\noindent More generally, let ${\cal S}(\varepsilon,n)$ be a function
\begin{itemize} 
\item which is for each integer $n$ with $n\geq\lambda$ analytic in $\varepsilon$ throughout an annular region centered by $0$; let ${\cal S}(\varepsilon,n)=\sum_{i=t}^{\infty}F_i(n)\varepsilon^i$ be its Laurent series expansion for some $t\in\set Z$;
\item  which satisfies (as $\ep$-expansion) the recurrence
\begin{equation*}
a_0(\varepsilon,n){\cal S}(\varepsilon,n)+\dots+a_d(\varepsilon,n){\cal S}(\varepsilon,n+d)=h_t(n)\ep^t+h_{t+1}(n)\ep^{t+1}\dots
\end{equation*}
for polynomials $a_i(\ep,n)\in\set K[\ep,t]$ with $a_d(0,n)\neq 0$ for all $n\geq\lambda$ and for functions $h_i(n)$ where the first coefficients $h_t(n),h_{t+1}(n)\dots,h_u(n)$ with $n\geq\lambda$ are expressible in terms of indefinite nested sums and products. 
\end{itemize}

\noindent Then there is the following algorithm~\cite[Cor.~1]{BKSF:12} implemented in \SigmaP:\\
\noindent\textbf{Input:} the polynomials $a_i(\ep,n)$ and the product-sum expressions $h_i(n)$ ($t\leq i\leq u$) as above; the values $c_{ij}$ ($t\leq i\leq u$, $\lambda\leq j\leq \lambda+d)$ such that 
${\cal S}(\ep,j)=c_{tj}\ep^t+c_{t+1,j}\ep^{t+1}+\dots+c_{u,j}\ep^{u}+\dots$\\
\textbf{Output:} the maximal $r\in\{-\infty,t,t+1,\dots,u\}$ such that the coefficients $F_t(n),F_{t+1}(n)\dots,F_r(n)$ of the $\ep$-expansion of ${\cal S}(\varepsilon,n)$ can be expressed in terms of indefinite nested sums and products. If $r\neq-\infty$, these representations of the coefficients are computed explicitly.

\medskip

\noindent For rigorous proofs, further details concerning efficiency, generalizations, and the function call within the package Sigma we refer to~\cite{BKSF:12}.

\section{Calculating $\ep$-expansions for multi-integrals}\label{Sec:Integrate}

We aim at computing a recurrence relation for multi-integrals of the form~\eqref{Equ:HypInt} and finding a Laurent-series solution that agrees with the input integral.
Here we exploit an enhanced version~\cite{Ablinger:12} of the multivariate Almkvist--Zeilberger algorithm~\cite{ZEILB} which contains as input class these integrals. More generally, it can handle integrands being hyperexponential in the integration variables $x_i$ (i.e., the logarithmic derivative of the integrand w.r.t.\ $x_i$ is a rational function in the $x_i$ and $n$) and hypergeometric in the discrete parameter $n$ (i.e., the shift quotient w.r.t.\ $n$ of the integrand is a rational function in the $x_i$ and $n$).

In order to illustrate the basic ideas, consider the double integral  
\begin{equation}\label{Equ:DoubleInt}
{\cal I}(\ep,n)=\int_0^1\int_0^1\underbrace{\frac{(1+x_1\cdot x_2)^n}{(1+x_1)^\ep}}_{F(n,x_1,x_2):=}dx_1dx_2
\stackrel{?}{=}F_0(n)+F_1(n)\ep+F_2(n)\ep^2+\dots
\end{equation}
First, one applies the multivariate Almkvist--Zeilberger algorithm. To be more precise, given $d\in\set N$ one looks for polynomials $e_i(n)$ and rational functions $R_i(n,x_1,x_2)$ such that
\begin{multline*}
e_0(n)F(n,x_1,x_2)+e_2(n)F(n+1,x_1,x_2)+\dots+e_d(n)F(n+d,x_1,x_2)\\
=D_{x1}(R_1 F(n,x_1,x_2))+D_{x2}(R_2 F(n,x_1,x_2));
\end{multline*}
here $D_{x_i}$ stands for the differentiation w.r.t.\ $x_i$. Internally, a clever ansatz is performed with undetermined coefficients which amounts to solving a linear system of equations. To hunt for a solution, one starts with $d=0$ and increases the recurrence order $d$ step by step until a solution is found. In our particular case, for $d=1$, one gets
\begin{equation}\label{Equ:IntegrandRec}
-(n+1)F(n,x_1,x_2)+(n+2)F(n+1,x_1,x_2)=D_{x_1}0+D_{x_2}\frac{x_2(x_1\cdot x_2+1)^{n+1}}{(1+x_1)^{\ep}}.
\end{equation}
Applying now the two integral signs on both sides of~\eqref{Equ:IntegrandRec} leads to the following recurrence
\begin{equation}\label{mAZExrec1}
-(n+1){\cal I}(\ep,n)+(n+2){\cal I}(\ep,n+1)=\underbrace{\int_0^1(x_1+1)^{n+1-\ep}dx_1}_{I_1(\ep,n)}-\int_0^1 0 dx_1. 
\end{equation}
Note that the right hand side of~\eqref{Equ:IntegrandRec} consists again of an integral. However, the nested depth is decreased from two to one. By recursion, we treat now this simpler integral again by the method under consideration. In this case, we find
\begin{equation*}
I_1(\ep,n)=\tfrac{2^{n+2}-1}{n+2}+\tfrac{\left(-2^{n+2} \left(\log(2) (n+2)-1\right)-1\right)}{(n+2)^2}\ep\\
	    +\tfrac{\left(2^{n+1}\left(\log(2)^2 (n+2)^2-2 \log(2) (n+2)+2\right)-1\right)}{(n+2)^3}\ep^2+\dots
\end{equation*}
Finally, plugging this result into \eqref{mAZExrec1} gives a recurrence that fits into the input class of the recurrence solver presented in Section~\ref{Sec:LaurentSolver}. Together with the expanded initial values (which can be calculated easily)
$${\cal I}(\ep,0)=\tfrac{5}{4}+\left(\tfrac{7}{8}-2 \log(2)\right)\ep+ \left(\tfrac{11}{16}+\log^2(2)-\tfrac{3 \log (2)}{2}\right)\ep^2+\dots$$
we are in the position to calculate the first three coefficients $F_i(n)$ of the expansion~\eqref{Equ:DoubleInt}.

In order to calculate this integral within Mathematica, several packages have to be load in: the \SigmaP\ package to use the recurrence solver from Section~\ref{Sec:LaurentSolver}, and the {\tt EvaluateMultiSum} package to deal with expansions. Finally, we load in the package {\tt MultiIntegrate} which contains an efficient implementation of the multivariate Almkquist--Zeilberger algorithm with the help of homomorphic image testing; see~\cite{Ablinger:12}. In addition, it combines all the steps described above (together with variations of the presented method) to perform the $\ep$-expansion.

\begin{mma}
\In << Sigma.m \vspace*{-0.06cm}\\
\Print Sigma - A summation package by Carsten Schneider
\copyright\ RISC\\
\In << EvaluateMultiSums.m \vspace*{-0.06cm}\\
\Print EvaluateMultiSums by Carsten Schneider -- \copyright\ RISC\\
\In << MultiIntegrate.m \vspace*{-0.06cm}\\
\Print MultiIntegrate by Jakob Ablinger -- \copyright\ RISC\\
\end{mma}

\noindent Now we are ready to calculate the expansion of the integral above. The coefficients are returned in list form, i.e., $\{F_0(n),F_1(n),F_2(n)\}$:

\begin{mma}
\In {\text{\bf sol=mAZExpandedIntegrate$\Big[\frac{(1 + x_1\; x_2)^n}{(1 + x_1)^\ep}$, n, \{$\ep$, 0, 2\}, \{\{$x_1$, 0, 1\}, \{$x_2$, 0, 1\}\}$\Big]$}}\vspace*{-0.3cm}\\
\Out \big\{\frac{1}{(n+1)}\Big(2 \sum_{i_1=1}^n
\tfrac{2^{i_1}}{1+i_1}-\sum_{i_1=1}^n \tfrac{1}{1+i_1}+1\Big),\frac{1}{(n+1)}\Big(1-2 \text{ln2}\,\Big(\sum_{i_1=1}^n
\tfrac{2^{i_1}}{1+i_1}+1\Big)-\sum_{i_1=1}^n
\tfrac{1}{\big(1+i_1\big){}^2}+2 \sum_{i_1=1}^n
\tfrac{2^{i_1}}{\big(1+i_1\big){}^2}\Big),\newline
\frac{1}{(n+1)}\Big(\text{ln2}^2\,
\Big(\sum_{i_1=1}^n \tfrac{2^{i_1}}{1+i_1}+1\Big)-2 \text{ln2}\,
\Big(\sum_{i_1=1}^n 
\tfrac{2^{i_1}}{\big(1+i_1\big){}^2}+1\Big)-\sum_{i_1=1}^n
\tfrac{1}{\big(1+i_1\big){}^3}+2 \sum_{i_1=1}^n
\tfrac{2^{i_1}}{\big(1+i_1\big){}^3}+1\Big)\big\}\\
\end{mma}

\noindent Note that the involved sums can be rewritten in terms of $S$-sums, see~\eqref{Equ:SSums}, using the command \texttt{TransformToSSums} from the package \texttt{HarmonicSums}:

\begin{mma}
\In << HarmonicSums.m \vspace*{-0.06cm}\\
\Print HarmonicSums by Jakob Ablinger -- \copyright\ RISC\\
\In TransformToSSums[sol]\vspace*{-0.3cm}\\
\Out \big\{\frac{S_1(2,n)}{n+1}-\frac{S_1(n)}{n+1}+\frac{2^{n+1}}{(n+1)^2}-\
\frac{1}{(n+1)^2},\text{ln2}\,
\Big(\frac{-S_1(2,n)}{n+1}-\frac{2^{n+1}}{(n+1)^2}\Big)+\frac{S_2(2,n)}{n+1}-\frac{S_2(n)}{n+1}+\frac{(2^{n+1}-1)}{(n+1)^3},\newline
\text{ln2}^2\, \Big(\frac{S_1(2,n)}{2 (n+1)}+\frac{2^n}{(n+1)^2}\Big)+\text{ln2} 
\,\Big(-\frac{S_2(2,n)}{n+1}-\frac{2^{n+1}}{(n+1)^3}\Big)+\frac{S_3(2,n)
}{n+1}-\frac{S_3(n)}{n+1}+\frac{2^{n+1}}{(n+1)^4}-\frac{1}{(n+1)^4}\big\}\\
\end{mma}

\noindent The proposed method extends to integrals with higher nesting depth. We conclude this approach by the calculation the coefficients $F_0(n),F_1(n),F_2(n)$ of the $\ep$-expansion of the following triple integral:
$$\int_0^1\int_0^1\int_0^1\frac{x_1^{\ep}(x_1+x_2x_3)^n}{ 
(1+x_3)^{\ep}}dx_1dx_2dx_3=F_0(n)+F_1(n)\ep+F_2(n)\ep^2+\dots$$
Integrals of this type emerge as partial integrals, e.g., for 3-loop ladder 
topologies.

\begin{mma}
\In mAZExpandedIntegrate[\tfrac{x_1^{\ep}(x_1+x_2x_3)^n}{ (1+x_3)^{\ep}},n,\{\ep,0,2\},\vspace*{-0.cm}\newline
\hspace*{1cm}\{\{x_1,0,1\},\{x_2,0,1\},\{x_3,0,1\}\}]//TransformToSSums//ReduceToBasis\vspace*{-0.4cm}\\
\Out \big\{\tfrac{-3 n-4}{(n+1)^2 (n+2)^2}+\tfrac{2^{n+1} (3 n+4)}{(n+1)^2 (n+2)^2}-\tfrac{S_1(n)}{(n+1) (n+2)}+\tfrac{S_1(2,n)}{(n+1) (n+2)},\quad
-\tfrac{2^{n+1} \big(3 n^2+6 n+2\big)}{(n+1)^3 (n+2)^3}+\tfrac{4 n^2+9 n+4}{(n+1)^3 (n+2)^3}+\tfrac{2 S_1(n)}{(n+1) (n+2)^2}-\tfrac{S_2(n)}{(n+1) (n+2)}-\tfrac{(-1)^n S_{-1}(n)+2 S_1(2,n)}{(n+1) (n+2)^2}+\text{ln2} \big(-\tfrac{2^{n+1} (3 n+4)}{(n+1)^2 (n+2)^2}-\tfrac{S_1(2,n)}{(n+1) (n+2)}-\tfrac{(-1)^n}{(n+1) (n+2)^2}+\tfrac{1}{(n+1) (n+2)^2}\big)+\tfrac{S_2(2,n)}{(n+1) (n+2)},
\newline
\big(\tfrac{2^n (3 n+4)}{(n+1)^2 (n+2)^2}+\tfrac{S_1(2,n)}{2 (n+1) (n+2)}+\tfrac{(-1)^n}{2 (n+1) (n+2)^2}-\tfrac{1}{2 (n+1) (n+2)^2}\big) \text{ln2}^2+\big(\tfrac{-3 n-5}{(n+1)^2 (n+2)^3}
\tfrac{2^{n+1} \big(3 n^2+6 n+2\big)}{(n+1)^3 (n+2)^3}-\tfrac{S_1(n)}{(n+1) (n+2)^2}+\vspace*{-0.3cm}\newline
+(-1)^n \big(\tfrac{S_{-1}(n)}{(n+1) (n+2)^2}+\tfrac{1}{(n+1) (n+2)^3}\big)+\tfrac{2 S_1(2,n)}{(n+1) (n+2)^2}-\tfrac{S_2(2,n)}{(n+1) (n+2)}\big) \text{ln2}-\tfrac{S_1(n){}^2}{2 (n+1) (n+2)^2}-\tfrac{(3 n+4) \big(3 n^2+9 n+7\big)}{(n+1)^4 (n+2)^4}\newline
+\tfrac{2^{n+1} \big(7 n^3+30 n^2+44 n+22\big)}{(n+1)^4 (n+2)^4}-\tfrac{2 (2 n+3) S_1(n)}{(n+1)^2 (n+2)^3}+\tfrac{S_2(n)-S_2(2,n)}{2 (n+1) (n+2)^2}-\tfrac{S_3(n)}{(n+1) (n+2)}+\tfrac{2 (2 n+3) S_1(2,n)}{(n+1)^2 (n+2)^3}+\tfrac{S_3(2,n)}{(n+1) (n+2)}\newline
+\tfrac{S_{1,1}(1,2,n)}{(n+1) (n+2)^2}+(-1)^n \big(\tfrac{S_{-1}(n)}{(n+1) (n+2)^3}+\tfrac{2 S_{-2}(n)}{(n+1) (n+2)^2}+\tfrac{S_{-1,1}(n)}{(n+1) (n+2)^2}-\tfrac{S_2(-2,n)}{(n+1) (n+2)^2}-\tfrac{S_{1,1}(-1,2,n)}{(n+1) (n+2)^2}\big)\big\}\\
\end{mma}

\noindent Here the command \texttt{ReduceToBasis} ensures that no algebraic relations exist among the arising $S$-sums; see also~\cite{Bluemlein:04,ANCONT3,ABC:11,Ablinger:12,ABC:12}.

\section{Calculating $\ep$-expansions for multi-sums}\label{Sec:Sum}

In our second strategy we rely on a common summation framework~\cite{Schneider:05b} of the difference field and holonomic approach~\cite{ZEILB,Chyzak}. The input class of the proposed method covers multi-sums of the form~\eqref{Eq:GenericMultiSum}. More generally, the summand itself can be an expression in terms of indefinite nested sums and products.
To illustrate this approach we aim at calculating the first coefficients of the double sum
$$\mathcal{S}(\ep,n)=\sum_{k=0}^{n-3}\underbrace{\sum_{j=1}^{n-k-2} \tfrac{(-1)^j \binom{-k+n-2}{j}\Gamma (j+k+1) \Gamma (n-k) 
\big(1-\frac{\ep}{2}\big)_k \big(2-\frac{\ep}{2}\big)_j}{\
\Gamma (-k+n-1) (3-\ep)_{j+k} \big(\frac{\ep}{2}+3\big)_{j+k}}}_{F(n,k)}
\stackrel{?}{=}F_0(n)+F_1(n)\ep+F_2(n)\ep^2+\dots$$
First one considers the inner sum $F(n,k)$ which itself is an analytic function in $\ep$ for each integer $n$ and $k$ with $n\geq3$ and $0\leq k\leq n-3$. First, we hunt for a recurrence relation in $k$. As described in Section~\ref{Sec:LaurentSolver} one can use the package Sigma and calculates
\begin{multline}\label{Equ:PureRec}
(k+1) (\ep-2 k-2)(k-n+2)F(n,k)\\
-(k-n+1) \big(\ep^2-\ep k+\ep n-\ep-4 k^2+2 k n-14 k-14\big)F(n,k+1)\\
-2 (\ep+k+2)(k-n+1) (k-n+2)F(n,k+2)=e_0(n,k)+e_1(n,k)\ep+e_2(n,k)\ep^2+\dots
\end{multline}
with
\scriptsize
\begin{align*}
e_0&(n,k)=-\frac{16 (k-n+1) (k-n+2) \big(k^2 n+4 k^2+2 k n+22 k-4 n+28\big)}{(k+2)^2 (k+3)^2 (k+4)^2},\\
e_1&(n,k)=\frac{8(k-n+1) (k-n+2) \big(2 k^5 n+10 k^5+18 k^4 n+134 k^4+43 k^3 n+698 k^3-32 k^2 n+1760 k^2-220 k n+2140 k-184 n+1000\big)}{(k+2)^3 (k+3)^3 (k+4)^3},\\
e_2&(n,k)=
-\frac{1}{2}(k+1) k! (k-n+1) (k-n+2) \big(\frac{16 \big(k^2 n+4 k^2+2 k n+22 k-4 n+28\big) S_2(k)}{(k+1) (k+2)^2 (k+3)^2 (k+4)^2 k!}\\
&-2 \Big(5 k^{10} n+4 k^{10}+142 k^9 n+230 k^9+1602 k^8 n+3812 k^8+9552 k^7 n+31332 k^7+32861 k^6 n\\
&\quad+150820 k^6+64970 k^5 n+452886 k^5+64724 k^4 n+857364 k^4+11424 k^3 n+985264 k^3-29328 k^2 n\\
&\quad+606320 k^2-11712 k n+121344 k+5952 n-30144\Big)\Big/({(k+1)^3 (k+2)^4 (k+3)^4 (k+4)^4 k!})\big).
\end{align*}
\normalsize
In addition, one computes a mixed recurrence, i.e., besides shifts in $k$ we allow in addition one shift in $n$, but keep $k$ unchanged. Using again Sigma, one obtains
\begin{multline}\label{Equ:OnePureRec}
(k-n) \big(-\ep^2+\ep k-\ep n+2 k^2-2 k n+2 k+2 n^2+2 n+2\big)F(n,k)\\
+(\ep-n-1) (\ep+2 n+2) (k-n+1) F(n+1,k)\\
-2 (\ep+k+1)(k-n) (k-n+1)F(n,k+1)=f_0(n,k)+f_1(n,k)\ep+f_2(n,k)\ep^2+\dots
\end{multline}
with
\scriptsize
\begin{align*}
f_0&(n,k)=\frac{16 (k-n) (k-n+1) \big(k^2-2 k n+2 k-4 n-1\big)}{(k+1) (k+2)^2 (k+3)^2},\\
f_1&(n,k)=-\frac{8(k-n) (k-n+1) \big(2 k^5-4 k^4 n+12 k^4-26 k^3 n+17 k^3-56 k^2 n-17 k^2-42 k n-47 k-4 n-19\big)}{(k+1)^2 (k+2)^3 (k+3)^3},\\
f_2&(n,k)=-\frac{1}{2} (k+1) k! (k-n) (k-n+1) \Big(-\frac{16 \big(k^2-2 k n+2 k-4 n-1\big) S_2(k)}{(k+1)^2 (k+2)^2 (k+3)^2 k!}\\
&\;\;+2 \Big(5 k^8-10 k^7 n+102 k^7-188 k^6 n+745 k^6-1300 k^5 n+2644 k^5-4456 k^4 n+4855 k^4-8162 k^3 n\\
&\;\;+4150 k^3-7644 k^2 n+635 k^2-2816 k n-944 k+96 n-192\Big)\Big/\Big((k+1)^4 (k+2)^4 (k+3)^4 k!\Big)\Big).
\end{align*}
\normalsize
We emphasize
that these two recurrences together with the initial values $F(4,0)=-\frac{13}{12}+\frac{23\ep}{288}-\frac{29\ep^2}{768}\dots$ and $F(4,1)=-\frac{1}{18}+\frac{11\ep}{432}-\frac{11\ep^2}{3456}
+\dots$ enables one to calculate the first three coefficients of the $\ep$-expansion for each $F(n,k)$ with $n\geq4$ and $0\leq k\leq n-3$. 

Given the two recurrences above (with this particular shape of shifts), one can apply the algorithm from~\cite{Schneider:05b} to calculate a recurrence for $\mathcal{S}(\ep,n)$. Using again Sigma, one gets the relation
\begin{multline}\label{Equ:HoloFinalRec}
\ep (\ep+2)^2 (2 n+1) (\ep-n-3) (\ep-n-2) (\ep+2 n+2) (\ep-n-1)^3 \mathcal{S}(\ep,n+1)\\
+4 \ep(\ep+2) (n+1)^3 (n+3) (2 n+3) (\ep-n-2) (\ep-n-1)\mathcal{S}(\ep,n)\\
=h_0(n)+h_1(n)\ep+h_2(n)\ep^2+h_3(n)F(n,0)\\
+h_4(n)F(n,1)+h_5(n)F(n,n-3)+h_6(n)F(n,n-2)+\dots
\end{multline}

\vspace*{-0.3cm}

\noindent with

\vspace*{-0.2cm}

\scriptsize
\begin{align*}
h_0(n)=&-\frac{64 \big(15 n^9+5 n^8-188 n^7-194 n^6+655 n^5+1129 n^4-266 n^3-1660 n^2-1224 n-288\big)}{9 (n-1)^2 n^2},\\
h_1(n)=&=\tfrac{32\big(15 n^{13}+87 n^{12}-45 n^{11}-329 n^{10}-215 n^9+249 n^8+2331 n^7+1545 n^6-6396 n^5-5440 n^4+6686 n^3+7776 n^2+648 n-864\big)}{27 (n-1)^3 n^3},\\
h_2(n) =&\frac{16 \big(4 n^{12}+16 n^{11}-23 n^{10}-188 n^9-219 n^8+208 n^7+503 n^6+12 n^5-445 n^4-96 n^3+340 n^2+272 n+64\big) S_2(n)}{(n-1)^2 n^2}\\
&\quad-\Big(4(708 n^{16}+916 n^{15}-9426 n^{14}-15711 n^{13}+30903 n^{12}+61829 n^{11}-49101 n^{10}-115105 n^9+44049 n^8\\
&\quad\quad+121347 n^7-789 n^6-57380 n^5-41832 n^4-14040 n^3+16848 n^2+6048 n-3456\big)\Big)\Big/\big(27 (n-1)^4 n^4\big),\\
h_3(n)=& \big(-\tfrac{4 \ep^2 \big(2 n^6+3 n^5-14 n^4-9 n^3-49 n^2-115 n-62\big) (n+1)^2}{n-1}+\tfrac{8 \ep \big(2 n^4+10 n^3-15 n^2-30 n+8\big) (n+1)^3}{n-1}-\tfrac{32 \big(3 n^2+10 n+8\big) (n+1)^3}{n-1}\big),\\
h_4(n)=&-32 (3 n+4) (n+1)^3+8\big(2 n^3-6 n^2-25 n-12\big) (n+1)^3\ep-4\big(2 n^5-3 n^4+16 n^3+32 n^2-18 n-31\big)(n+1)^2\ep^2,\\
h_5(n)=&16 (n-2)^2 \big(3 n^3+n^2-7 n-4\big) (n+1)^3-4\big(2 n^7-15 n^6+46 n^5-18 n^4-125 n^3+86 n^2+96 n-16\big) (n+1)^2\ep\\
&\quad+2\big(16 n^8-58 n^7-7 n^6+221 n^5-59 n^4-325 n^3-26 n^2+146 n+60\big)\ep^2,\\
h_6(n)=&32 \big(3 n^4-8 n^3-10 n^2+20 n+16\big)(n+1)^3-8\big(2 n^6-13 n^5+27 n^4+56 n^3+14 n^2+66 n+60\big)(n+1)^2\ep\\
&\quad+8\big(6 n^7-23 n^6-43 n^5+46 n^4+68 n^3-23 n^2-43 n-12\big)\ep^2;
\end{align*}
\normalsize
for the different command calls within \SigmaP\ we refer to~\cite{Schneider:05b,Schneider:07a}.
As indicated above, the algorithm itself only uses the two recurrence relations~\eqref{Equ:PureRec} and~\eqref{Equ:OnePureRec} and thus the occurring expressions $F(n,0),F(n,1),F(n,n-3),F(n,n-2)$ remain unevaluated.
Next, one applies the proposed method recursive on these sums.
Since these objects are simpler than the input sum $\mathcal{S}(\ep,n)$, the termination of our method is guaranteed. E.g., the calculation of the $\ep$-expansion
\begin{align*}
F(n,0)&=\sum_{j=1}^{n-2}\tfrac{(-1)^j \Gamma (j+1) \Gamma (n)
   \binom{n-2}{j}
   \left(2-\frac{\ep}{2}\right)_j}{\Gamma (n-1) (3-\ep)_j
   \left(\frac{\ep}{2}+3\right)_j}\\
&=4 (n-1) \big(\tfrac{5-n}{4 (n-1)}-\tfrac{S_1(n)}{(n-1) n}\big)+ (1-n) \big(\tfrac{S_1(n){}^2}{(n-1) n}-\tfrac{S_1(n)}{(n-1) n}+\tfrac{S_2(n)}{(n-1) n}+\tfrac{1}{1-n}\big)\ep\\ 
&+\tfrac{n-1}{6}\big(-\tfrac{S_1(n){}^3}{(n-1) n}+\tfrac{3 S_1(n){}^2}{2 (n-1) n}+\big(\tfrac{15}{(n-1) n}
-\tfrac{3 S_2(n)}{(n-1) n}\big) S_1(n)+\tfrac{3 S_2(n)}{2 (n-1) n}-\tfrac{2 S_3(n)}{(n-1) n}-\tfrac{12 S_{2,1}(n)}{(n-1) n}\big)\ep^2+\dots
\end{align*}
boils down to the method described in Section~\ref{Sec:LaurentSolver}.
Similarly one proceeds for $F(n,1)$ and $F(n,n-3),F(n,n-2)$. This finally leads to the following simplified right and side of~\eqref{Equ:HoloFinalRec}:
\small
\begin{align*}
&0\times\ep^0 + \Big(32 n (n+1)^2 (3 n+4) \big(n^2+3 n+4\big)-128 (n+1)^3 (3 n+4) S_1(n)\Big)\ep\\
&+\Big(-32 \big(2 n^2+7 n-2\big) (n+1)^3 S_1(n)+16 \big(4 n^5+12 n^4-27 n^3-129 n^2-130 n-28\big) (n+1)^3 S_2(n)\\
&-8 n \big(12 n^6+58 n^5-27 n^4-566 n^3-1125 n^2-804 n-156\big) (n+1)-32 (3 n+4) (n+1)^3 S_1(n){}^2\Big)\ep^2+\dots
\end{align*}
\normalsize
Note that the left hand side of~\eqref{Equ:HoloFinalRec} and its right hand side (after the simplification) can be divided by $\ep$, i.e., the coefficient of $\mathcal{S}(\ep,n+1)$ evaluated at $\ep=0$ does not vanish. Hence together with the expanded initial values
$\mathcal{S}(\ep,3)= -\frac{4}{9}+\frac{1}{27}\ep+\dots$ and $\mathcal{S}(\ep,4)=-\frac{41}{36} +\frac{91}{864}\ep+\dots$ one can activate our recurrence solver for Laurent series to calculate the first two coefficients of the $\ep$-expansion 
\begin{align*}
\mathcal{S}(\ep,n)=&-\tfrac{8}{(n+1) (n+2)}S_1(n)-4 (2 n+1) S_2(n)+\tfrac{4 n (3 n+7)}{n+2}
+\Big(\tfrac{-2 \big(2 n^3+3 n^2+3 n+6\big)}{(n+1)^2 (n+2)^2}S_1(n)\\
&+\tfrac{\big(6 n^3+23 n^2+27 n+12\big)}{(n+1) (n+2)}S_2(n)
-\tfrac{2}{(n+1) (n+2)}S_1(n){}^2-2 (2 n+1) S_3(n)-\tfrac{n (n+1) (5 n+14)}{(n+2)^2}\Big)\ep+\dots
\end{align*}

\noindent Summarizing, in the presented method one constructs step by step suitable inhomogeneous recurrences from the innermost sum to the outermost sum\footnote{
In~\cite{Chyzak} this idea has been considered for homogeneous recurrences with polynomial coefficients. In our approach~\cite{Schneider:05b} we observed that setting up the recurrence system in the special form given above (instead of allowing a general holonomic system) one can derive an efficient algorithm without using Gr\"obner basis. In this way, the holonomic approach could be extended in~\cite{Schneider:05b} to handle also inhomogeneous recurrences formulated in difference fields. In order to take into account the $\ep$-expansion of the inhomogeneous sides, new ideas have been added into \SigmaP.}.
As one can see already for double sums, this construction is quite involved and is fairly complicated for more nested sums (e.g., taking care of poles, estimating how far one should expand\footnote{E.g., in the illustrated example from above one has to start to calculate three coefficients of the $\ep$-expansion and ends up only with the first two coefficients.}, or exploiting a refined difference field theory~\cite{Schneider:08c}). 
The new package \texttt{RhoSum} deals with all these aspects using as backbone the packages \texttt{Sigma}, \texttt{HarmonicSums}, and \texttt{EvaluateMultiSums}. After loading

\begin{mma}
\In << RhoSum.m \vspace*{-0.06cm}\\
\Print RhoSum -  Package for Refined Holonomic Summation
\copyright\ RISC\\
\end{mma}

\noindent we can perform the calculation from above with the function call

\begin{mma}
\In FindSum[\tfrac{(-1)^j \binom{-k+n-2}{j}\text{$\Gamma$}(j+k+1) \text{$\Gamma$}(n-k) 
\big(1-\frac{\ep}{2}\big)_k \big(2-\frac{\ep}{2}\big)_j}{
\text{$\Gamma$}(-k+n-1) (3-\ep)_{j+k} \big(\frac{\ep}{2}+3\big)_{j+k}}
,\{\{j,1,n-k-2\},\{k,0,n-3\}\},\{n\},\{3\},\{\infty\},\vspace*{-0.2cm}\newline
\hspace*{10cm}ExpandIn\to\{ep,0,1\}\vspace*{-0.5cm}\\
\Out\big\{-\tfrac{8}{(n+1) (n+2)}S_1(n)-4 (2 n+1) S_2(n)+\tfrac{4 n (3 n+7)}{n+2},\newline
\tfrac{-2 \big(2 n^3+3 n^2+3 n+6\big)}{(n+1)^2 (n+2)^2}S_1(n)
+\tfrac{\big(6 n^3+23 n^2+27 n+12\big)}{(n+1) (n+2)}S_2(n)
-\tfrac{2}{(n+1) (n+2)}S_1(n){}^2-2 (2 n+1) S_3(n)-\tfrac{n (n+1) (5 n+14)}{(n+2)^2}\big\}\\
\end{mma}

\noindent A more involved problem is, e.g.,

\begin{multline*}
\sum_{j=0}^{n-2}\sum_{j_1=0}^j\sum_{j_2=1}^{n-4+j_1-j}
\tfrac{(-1)^{-{j_1}-{j_2}+n-4} e^{-\frac{3 \ep \gamma }{2}} \binom{j}{{j_1}} \binom{-j+{j_1}+n-4}{{j_2}}\Gamma\big(-\frac{\ep}{2}-j+{j_1}-{j_2}+n-1\big)\Gamma(-\ep-j+n-1) \Gamma\big(-\frac{\ep}{2}-j+{j_1}+n\big)}{(\ep+1) (\ep+2) \Gamma\big(2-\frac{\ep}{2}\big) \Gamma(\ep+4) \Gamma(j-{j_1}+{j_2}+2) \Gamma(-2 \ep-j+{j_1}+n)}\times\\
\times\tfrac{\Gamma\big(\frac{\ep}{2}+3\big)^2 \Gamma\big(-\frac{3 \ep}{2}\big) \Gamma\big(-\frac{\ep}{2}\big) \Gamma(\ep) \Gamma(j-{j_1}+1) \Gamma(-\ep+{j_1}+1) \Gamma({j_2}+1)}{\Gamma\big(\frac{\ep}{2}-j+{j_1}+n\big) \Gamma\big(-\frac{\ep}{2}-j+{j_1}-{j_2}+n\big)}\stackrel{?}{=}F_{-3}\ep^{-3}+F_{-2}\ep^{-2}+F_{-1}\ep^{-1}+F_{0}+\dots
\end{multline*}
Sums of this type occur, e.g., in case of 3-loop topologies with one massive and one 
massless fermion line.
If we insert this sum into Mathematica in the variable $f$, then we get the following expansion. The constant term is too large to present it here.

\begin{mma}
\In FindSum[f
,\{\{j_2,1,n-4+j_1-j\},\{j_1,0,j\},\{j,0,n-2\}\},\{n\},\{2\},\{\infty\},ExpandIn\to\{ep,-3,-1\}\\
\Out \big\{-\tfrac{4 \big(n^3-5 n^2+6 n-4\big)}{9 (n-2)^2 (n-1) n^3}+\tfrac{2 (-1)^n \big(n^4-2 n^3+n^2-12 n+8\big)}{9 (n-2) (n-1) n^3}-\tfrac{8 (-1)^n (n-3) S_ 1(n)}{9 (n-2) n^2},\newline 
+\tfrac{16 (-1)^n S_ {-2}(n)}{9 (n-2) n^2}+ \tfrac{(-1)^n (10-n) S_ 1(n){}^2}{9 (n-2) n^2}+\big(\tfrac{(-1)^n \big(17 n^3-179 n^2+240 n-96\big)}{27 (n-2) (n-1) n^3}+\tfrac{4}{9 (n-2)^2 (n-1) n^2}\big) S_ 1(n)\newline
+\tfrac{(-1)^n \big(-2 n^5+19 n^4-5 n^3+111 n^2-190 n+84\big)}{27 (n-2) (n-1) n^4}+\tfrac{2 \big(8 n^5-63 n^4+163 n^3-218 n^2+140 n-36\big)}{27 (n-2)^2 (n-1)^2 n^4}+\tfrac{(-1)^n (14-n) S_ 2(n)}{9 (n-2) n^2},\newline 
 \tfrac{(-1)^n (22-n) S_ 1(n){}^3}{54 (n-2) n^2}+\big(\tfrac{(-1)^n \big(n^3-299 n^2+412 n-168\big)}{108 (n-2) (n-1) n^3}+\tfrac{1}{9 (n-2)^2 (n-1) n^2}\big) S_ 1(n){}^2\newline
 +\big(\tfrac{-6 n^2+5 n-5}{27 (n-2)^2 (n-1)^2 n^2}+\tfrac{(-1)^n \big(-313 n^4+1708 n^3-2262 n^2+1740 n-648\big)}{162 (n-2) (n-1) n^4}+\tfrac{(-1)^n (7 n+46) S_ 2(n)}{18 (n-2) n^2}\big) S_ 1(n)\newline
+\tfrac{-289 n^6+2430 n^5-6803 n^4+10222 n^3-8536 n^2+4176 n-1008}{324 (n-2)^2 (n-1)^2 n^5}\newline
 +\tfrac{(-1)^n \big(475 n^7-1768 n^6+2423 n^5-6170 n^4+17912 n^3-21856 n^2+15744 n-5184\big)}{648 (n-2)^2 (n-1) n^5}-\tfrac{2 (-1)^n (11 n-18) S_ {2,1}(n)}{9 (n-2) n^2}\newline
+S_ {-2}(n) \big(\tfrac{16 (-1)^n S_ 1(n)}{9 (n-2) n}-\tfrac{4 (-1)^n \big(9 n^3+22 n^2-43 n+18\big)}{27 (n-2) (n-1) n^3}\big)-\tfrac{2 (-1)^n (3 n-16) S_ {-3}(n)}{9 (n-2) n^2}\newline
+\big(\tfrac{-6 n^3+30 n^2-35 n+24}{9 (n-2)^2 (n-1) n^3}+\tfrac{(-1)^n \big(n^4-641 n^3+1646 n^2-976 n+192\big)}{108 (n-2)^2 (n-1) n^3}\big) S_ 2(n)+\tfrac{(-1)^n (47 n-8) S_ 3(n)}{27 (n-2) n^2}
\newline
-\tfrac{4 (-1)^n S_ {-2,1}(n)}{3 (n-2) n}+\big(\tfrac{-n^3+5 n^2-6 n+4}{6 (n-2)^2 (n-1) n^3}+\tfrac{(-1)^n \big(n^4-2 n^3+n^2-12 n+8\big)}{12 (n-2) (n-1) n^3}+\tfrac{(-1)^n (3-n) S_ 1(n)}{3 (n-2) n^2}\big) \zeta_2\Big\}\\
\end{mma}

\noindent The above expressions can be analytically continued to complex values of the Mellin 
variable $n$ using relations given in \cite{Blumlein:2000hw,Blumlein:2005jg,
Blumlein:2009ta,ANCONT3}.

\section{Conclusion}

Massive Feynman integrals with operation insertion can be expressed in terms of multi-in\-tegrals and multi-sums over hypergeometric and hyperexponential functions.
We presented new methods to calculate the first coefficients of the $\ep$-expansion of such multi-sums and multi-integrals. Here the multivariate Almkvist-Zeilberger algorithm and the common framework of the holonomic and difference field algorithms have been enhanced to calculate recurrences. Then a  recurrence solver for Laurent series expansion is used to extract the all $n$ coefficients of the $\ep$-expansion. Besides of the usage of the Mathematica packages \SigmaP, \texttt{HarmonicSums} and \texttt{EvaluateMultiSums}, two new packages \texttt{MultiIntegrate} and \texttt{Rho} have been developed that can carry out these calculations in a completely automatic fashion.

\vspace*{2mm}
\noindent
{\bf Acknowledgment.}
This work has been supported in part by DFG Sonderforschungsbereich Transregio 9, 
Computergest\"utzte Theoretische Teilchenphysik, Austrian Science  Fund (FWF) 
grant P203477-N18, and EU Network {\sf LHCPhenoNet} PITN-GA-2010-264564. 


\end{document}